**In depth nano spectroscopic analysis on homogeneously switching double barrier memristive devices**

Julian Strobel[a], Mirko Hansen[b], Sven Dirkmann[c], Krishna Kanth Neelisetty[d], Martin Ziegler[b], Georg Haberfehlner[e], Radian Popescu[f], Gerald Kothleitner[e], Venkata Sai Kiran Chakravadhanula[d], Christian Kübel[d], Hermann Kohlstedt[b], Thomas Mussenbrock[g], and Lorenz Kienle[a,*]

**a:** Synthesis and Real Structure, University of Kiel, Faculty of Engineering,

Kaiserstr. 2, 24143 Kiel, Germany

just@tf.uni-kiel.de;, lk@tf.uni-kiel.de

**b**: Nanoelectronics, University of Kiel, Faculty of Engineering,

Kaiserstr. 2, 24143 Kiel, Germany

mha@tf.uni-kiel.de, maz@tf.uni-kiel.de, hko@tf.uni-kiel.de

**c:** Ruhr Universität Bochum, Faculty for Electrical Engineering and Information Technology,

Universitätsstr. 150, 44801 Bochum, Germany

sven.dirkmann@gmx.de, mussenbrock@gmail.com

**d:** Karlsruhe Institute of Technology, Institute of Nanotechnology,

Hermann-von-Helmholtz-Platz 1, 76344 Eggenstein-Leopoldshafen, Germany

cvskiran@kit.edu, krishna.neelisetty@kit.edu

**e:** Technical University Graz, Institute for Electron Microscopy and Nanoanalytics,

Steyrergasse 17, 8010 Graz, Austria

georg.haberfehlner@felmi-zfe.at, gerald.kothleitner@felmi-zfe.at

**f:** Karlsruhe Institute of Technology, Laboratory of Electron Microscopy

Engesserstr. 7, 76131 Karlsruhe, Germany

popescu@kit.edu

**g:** Brandenburg University of Technology, Electrodynamics and Physical Electronics

03046 Cottbus, Germany

mussenbrock@b-tu.de

*Corresponding author:

Lorenz Kienle, Kaiserstr. 2, 24143 Kiel, Germany

Email: lk@tf.uni-kiel.de; Phone: 0049 431 880 6196

**ABSTRACT**

Memristors based on a double barrier design have been analysed by various nano spectroscopic methods to unveil details about its microstructure and conduction mechanism. The device consists of an $AlO_x$ tunnel barrier and a $NbO_y$/Au Schottky barrier sandwiched between Nb bottom electrode and Au top electrode. As it was anticipated that the local chemical composition of the tunnel barrier, i.e. oxidation state of the metals as well as concentration and distribution of oxygen ions, have a major influence on electronic conduction, these factors were carefully analysed. A combined approach was chosen in order to reliably investigate electronic states of Nb and O by electron energy-loss spectroscopy as well as map elements whose transition edges exhibit a different energy range by energy-dispersive X-ray spectroscopy like Au and Al. The results conclusively demonstrate significant oxidation of the bottom electrode as well as a small oxygen vacancy concentration in the Al oxide tunnel barrier. Possible scenarios to explain this unexpected additional oxide layer are discussed and kinetic Monte Carlo simulations were applied in order to identify its influence on conduction mechanisms in the device. In light of the strong deviations between observed and originally sought layout, this study highlights the robustness in terms of structural deviations of the double barrier memristor device.

# 1. INTRODUCTION

The scientific interest in resistive switching devices increases drastically when in 2008 Strukov et al. claimed that the fourth lumped circuit element has been found. [1] It was theoretically predicted by Chua in the early 1970s. [2] This device also referred to as the memristor has opened the door to a vast number of new and promising applications. The applications of memristors span from computer memories (RRAM) over computational pattern recognition to neuromorphic engineering. [3–6]

Unlike the transistor which initiated the electronic revolution and which of course has seen manifold modifications since its introduction, there is no single common design-scheme for memristive devices. Furthermore, the physical mechanisms which are responsible for the drastic change of the resistance are diverse as well. A categorization can be made based on the resistive switching mechanism: in electrochemical metallization (ECM) cells a conductive filament forms from what is coined the active electrode towards the inactive electrode. The active electrode and its conductive filament consists of non-inert metals like Ag or Cu. Inactive electrodes in all these devices may consist of noble metals such as Au and Pt or from conductive compounds, e.g. TiN. [7–10] In valence change memory (VCM) cells transition metal oxides between the electrodes locally change their oxidation state by movement of charged defects, such as oxygen vacancies. This behaviour can occur in the form of homogeneous drift of charge carriers or filamentary switching. [11,12] Phase change memory (PCM) devices exhibit the formation of crystalline filaments in an amorphous matrix. This behavior can be observed in chalcogenides such as complex tellurides [13] and e.g. in amorphous TiOx where crystalline Magnéli filaments form. [14] Of course this categorization is not complete as other less common types of switching exist, and like any classification it is far from absolute; e.g. a phase change can – and often does – include a valence change. Among these approaches two-terminal tunnel junctions [15–17] represent a very promising approach as they might be directly compatible to CMOS technology and work with small voltages and currents, hence consume only little power.

An elegant realization of a memristive device which does not rely on the disadvantageous formation and deformation of filaments has been proposed by Hansen et al. [16] It consists of a double tunnel barrier with a variable tunnel resistance. (cf. **Figure 1**) The conduction mechanism and switching principle are matter of ongoing research. Electrical characterization and numerical simulations [18] which has been already performed prove the concept. However, a profound experimental investigation from the nanoanalytical point of view has not been reported yet; consequently, the memristive double barrier device is investigated based on a dedicated electron microscopic analysis. A particular experimental challenge is to investigate nanoscale chemical segregation phenomena in the very thin amorphous layers which exhibit thicknesses around 10 nm. The determination of such features is highly relevant, since it is expected that mobile oxygen ions in the Schottky and the tunnel barrier are responsible for the resistance change. In order to address these open questions the experimental results have been analyzed in conjunction with Monte Carlo simulations of the device.

## 2. EXPERIMENTAL

### 2.1. Device fabrication

A stack of Nb/Al-AlO$_x$/NbO$_y$/Au was deposited onto thermally oxidized Si-SiO$_2$ wafers. The multilayer devices were fabricated by standard optical lithographic means and DC magnetron sputtering onto 4-inch wafers. Detailed deposition parameters can be found in [16]. The targeted thicknesses were 6 nm (Al), 1 nm (AlO$_x$), 1.2 nm (NbO$_y$). Nb and Au serve as bottom and top electrode and their thicknesses are large compared to the Al and oxide films. A schematic of the targeted layout is depicted in **Figure 1 (b)**. Because the design of the memristive device is identical to superconducting Josephson junctions except for the lack of the second oxide layer (cf. **Figure 1 (c)**), it has been analyzed in this study to the effect of validating results, because it is a well-understood system under analysis for decades.

2.2. **TEM sample preparation**

Cross-sectional samples were prepared by focused ion beam (FIB) milling on a FEI *Helios NanoLab* as well as FEI *Strata 400 S* and thinned down to thicknesses of about 100-150 nm. Final thinning was done by treatment with low energy Ar ions in a Fischione *NanoMill* to a final thickness of around 50 – 80 nm to prohibit any preparation artifacts from the FIB's Ga ions. Josephson junction wafers were also available as large scale samples for which cross-sectional sample preparation by Precision Ion Polishing System (PIPS) was applicable. As PIPS is expected to be less prone to preparation induced artifacts, a comparative study between FIB and PIPS prepared samples was conducted. In order to substantiate results even further, FIB backside thinning was applied to exclude any directional thinning artifacts, e.g. oxidation of the metal layers.

2.3. **TEM investigations and data analysis**

As the overall thickness of the critical layers in the double barrier memristor (DBM) is well below 10 nm and information both on the structure and local chemistry is required transmission electron microscopy (TEM) analysis was conducted. The chemical information was retrieved by energy-dispersive X-ray spectroscopy (EDX) and electron energy-loss spectroscopy (EELS) in scanning TEM (STEM). High-angle annular dark field (HAADF)-STEM combined with EDX experiments were performed on a FEI Osiris ChemiSTEM microscope at 200 kV acceleration voltage, which is equipped with a Bruker Quantax system (XFlash detector) for EDX. Elemental maps of Al (Al-$K_\alpha$ line), Nb (Nb-$K_\alpha$) and O (O-$K_\alpha$ line) were recorded and used to investigate their distribution within multi-layers. The maps were analyzed by using the ESPRIT software (version 1.9) from Bruker. Using ESPRIT, element concentrations were calculated on the basis of a refined Kramers' law model, which includes corrections for detector absorption and background subtraction. For this purpose, standard-less quantification, i.e. by means of theoretical sensitivity factors, without thickness correction was applied.

Combined STEM-EELS and EDX experiments were carried out on a FEI *Titan³* (monochromator, XFEG, GIF Quantum energy analyzer, *SuperX* EDX detector). The energy resolution during EELS was continuously checked to be at least 0.3 eV. Real space micrographs were taken by annular dark field imaging also in STEM mode. Data processing of the EEL spectra was undertaken with Gatan Microscopy Suite 3 and HyperSpy [19]. The data was treated by principal component analysis to reduce noise and find all significant features.[20] Oxygen K spectra were fitted with a multiple linear least square (MLLS) fit to a reference spectrum in order to locate specific oxides. The references were extracted from the middle of the respective oxide layers (Nb and Al) and then fit to the rest of the EELS data creating a goodness of fit map.

### 2.4. **Anodization spectroscopy**

The anodization of thin films through the application of a constant direct current and the simultaneous measurement of the resulting voltage allows investigating layer properties (e.g. thickness and roughness). Anodization spectroscopy has been applied successfully to optimize the fabrication of high-quality low-temperature all-niobium Josephson tunnel junctions.[21]

The set-up for anodization spectroscopy consists of an electrolyte bath, an Au electrode (cathode), the sample of interest (anode), a current source and a computer for data acquisition.[22] The most common method is to record the measured voltage V over time and subsequently plot the time derivative dV/dt (anodization rate) versus V at room temperature. Different materials exhibit different anodization rates, leading to a characteristic anodization spectrum were the bulk materials and the interface between layers are easily distinguishable.

The low-cost set-up together with the possibility to explore interfacial properties on the nanometer level and the lack of time consuming sample preparation (e.g. annealing, patterning etc.) made anodization spectroscopy an established and attractive method.[23] Although not all features of anodization profiles are understood yet, the different anodization rates of different materials and the transition between two layers allows comparative studies. Therefore, we applied anodization spectroscopy to study memristive double barrier Nb/Nb$_x$O$_y$/Al$_2$O$_3$-Al/Nb junctions. The top electrode had to be replaced by Nb because of the noble character of Au making it unsuitable for anodization spectroscopy.

## 3. RESULTS

### 3.1. Structural and spectroscopic analysis of DBM devices

After ensuring that the preparation methods are generally suitable for this material system the analysis of the actual memristive device was conducted. A high resolution micrograph revealing the metallic crystalline and amorphous oxide regions is depicted in **Figure 2 (d)**. The indicated thickness of 10 nm of the amorphous area (dashed lines) containing three distinct oxides is remarkable as it is much thicker than expected and unusually thick for a functioning tunnel barrier. Subsequently, chemical information of the sample was extracted by STEM-EDX experiments. Here, it is clearly visible (cf. **Figure 2 (a)** and **(b)**) that the Nb bottom electrode exhibits oxidation reaching several nanometers in thickness (labelled NbO$_z$ in the following). Furthermore, no isolated metallic Al seems to be present as apparent from the lateral distribution of oxygen. The NbO$_y$ deposited on top of the AlO$_x$, is ca. 1 nm thick and thus closely corresponds to the thickness predicted from deposition parameters; the combined thickness of the three oxide layers is about 10 nm and corresponds with the amorphous region visible in the HRTEM micrograph. The surface roughness clearly visible between the lower NbO$_z$ and AlO$_x$ is also in line with expectations and is similar to that observed for the Josephson junction (cf. supplement). A slight overlap of the NbO$_z$ and AlO$_x$ is present and may be caused by two factors. Firstly, the surface roughness causes a superposition of Nb and Al layers along the beam direction thus signal mixing for $NbO_z$ and $AlO_x$ is produced from the same apparent lateral position. Secondly, a slight interdiffusion may be possible. However, taking into account the tiny dimensions of the layers it seems hardly possible to differentiate between these two cases.

The source of Nb bottom electrode (BE) oxidation is unknown. Samples have been prepared by FIB milling both from the front- as well as from the backside, with and without NanoMill treatment in order to test for a Ga-beam induced oxidation during FIB preparation. As samples from all preparation routes yielded qualitatively the same results in the TEM, a directional – i.e. beam induced – artefact can be excluded (see supplementary material). Furthermore, the congruency of results between FIB and PIPS prepared Josephson junction suggests that the FIB does not alter the thin $NbO_y$ or $AlO_x$ layers; PIPS preparation is expected to cause no artifacts due to the use of low energy Ar ions.

Additionally, STEM-EELS investigations were conducted using a narrow energy range containing the O-K edge (532 eV), hence, more detailed chemical information was available. The O-K edge was fitted to reference spectra from Al oxide and Nb oxide. The references were extracted according to the method described in section 2.3. Maps of the goodness of this fit are depicted in **Figure 3**, clearly identifying the locations where oxygen is present in the form of $AlO_x$ and $NbO_{y/z}$, conclusively proving the oxidation of the BE as well as the absence of metallic Al. Most importantly though, the fits allow determination of the true thickness of the tunnel barrier at its thinnest locations. Due to 3D surface roughness, the thinnest areas of the tunnel barrier are always obscured by thicker regions along the beam direction. However, the extent of Nb oxide as determined from the O-K edge, proves that the tunnel barrier is in fact only around 1 nm thick at the thinnest spots, allowing tunnelling as proposed by Hansen et al. The yellow lines in **Figure 3** (a) indicate these regions. As the $NbO_y$ Schottky barrier exhibits no significant surface roughness, the maximum surface roughness extent of the NbO$_z$ (at the location below the yellow lines) implies the minimum thickness of the tunnel barrier.

Aside from the lateral distribution of elements the near edge structure (ELNES) of the O-K edge was analyzed in more detail. The reference spectra, also used for MLLS fitting of the maps shown in **Figure 3**, are displayed in **Figure S1**. The AlO$_x$ spectra all generally resemble the Al$_2$O$_3$ spectrum known from literature, which is no surprise as Al oxide is basically always a stoichiometric oxide and no sub- or superoxide. However, a pre-peak feature of the O-K edge at 538.6 eV emerges, shifted about 6 eV to lower energies compared to the main edge. This feature appears to be characteristic since it was observed across all samples. A close up of this feature is shown in **Figure 4** along with two fit lines for the main edge and its pre-peak. It has been reported in literature that irradiation by the electron beam can cause damage to Al containing oxides and produce similar features.[24,25] However, these spectra were recorded on a monochromated instrument and the dwell times were only 100 ms, thus irradiation damage should be marginal. Furthermore, no timely evolution of this feature was observed, which is expected for beam induced processes. This makes it improbable for this feature to be induced during the investigation. The possibility of this pre-peak emerging due to the influence of either of the neighbouring Nb oxides [26] was also considered. However, no Nb-M edge, neither M4/5, M3/2 nor M1, was observed, which were all perfectly visible in the NbO$_{y/z}$ regions. Furthermore, the pre-peak feature is present throughout the entire AlO$_x$ layer and not only at the interface to neighbouring Nb containing layers, hence, this scenario is excluded for interpretation.

As Nigo et al. [27] described in their study, though, such a feature could be indicative of oxygen vacancies. The electronic structure of alumina [28] and the influence of oxygen vacancies [29–31] on it have been described extensively in the literature. Accordingly, the area ratio of pre-peak to main edge is characteristic for the concentration of oxygen vacancies. As described by Johnson and Pepper, mid-band gap states in Al$_2$O$_3$ belong to localized Al(3s) states, which usually form bonding O(2p)-Al(3s) states and are located in the valence band. [32] The lower boundary of the conduction band is formed by the Al(3p) orbitals, accordingly this corresponds to the main peak of the O-K edge. As Al(3p) is threefold degenerate, three times the number of the Al(3p) states compared to overall – i.e. bound – Al(3s) states exist. Under the approximation that oxygen 1s electrons are excited into Al(3s) and Al(3p) states with equal probability, the integrated EELS intensities scale linearly with the number of states $N_{Al(orb.)}$ described above. Accounting for the local symmetry of Al$_2$O$_3$ every oxygen ion is coordinated with four aluminum ions; one vacancy hence produces four localized Al(3s) states. The total concentration of oxygen vacancies $c_{vac}$ is therefore given by:

$$\frac{3}{4} \cdot \frac{N_{Al(3s),loc}}{N_{Al(3p)}} = \frac{1}{4} \frac{N_{Al(3s),loc}}{N_{Al(3s)}} = c_{vac} \qquad (1)$$

Where the ratio of localized Al(3s) to Al(3p) (highlighted in green) is the value experimentally determined from ELNES analysis (cf. **Figure 4**). Oxygen vacancy concentrations were calculated accordingly, yielding integrated EELS intensities $I_{Al(3s),loc}/I_{Al(3p)}$ between 2.2 % and 4.3 % over different samples which results in oxygen vacancy concentrations of 1.7 % to 3.2 %. Since literature on this kind of quantification is scarce and verification of this result with samples of known concentrations was never reported, the accuracy of these values can only be estimated. Especially the relative scattering probability into bonding Al(3p) and localized Al(3s) orbitals is unknown. However, as the oxide is completely amorphous, differences from anisotropy of the electronic structure can excluded. Furthermore, both orbitals are completely empty, hence Coulomb repulsion or effects from electron spin can be excluded.

Identification of Al and Nb oxidation state was done by comparison with O-K ELNES spectra from literature [33,34]. As expected, the O-K near edge structure in the AlO$_x$ region resembles that of Al$_2$O$_3$ and NbO$_z$ region resembles that of Nb(II) oxide the most. The O-K ELNES from the NbO$_y$ barrier could not be identified as a specific Nb oxide, a mixture between Nb(IV) and Nb(V) seems likely, though. (cf. **Figure 5**) Comparison of edge cross sections supports the ELNES comparison, as the O/Nb ratio is 25 % higher in the NbO$_y$ Schottky barrier oxide than in the oxidized bottom electrode. Thus the degree of oxidation is higher in NbO$_y$ with an expected oxidation state to be at least Nb(IV). This, too, underlines the stability of the Schottky barrier at the NbO$_y$/Au interface.

### 3.2. **Anodization spectroscopy**

The black curve in **Figure 6** shows a typical anodization spectrum of an Nb/Al$_2$O$_3$-Al/Nb Josephson junction.[35] The anodization front propagates from the surface of the upper Nb layer. When the anodization front reaches the Al$_2$O$_3$ tunnel barrier at around 10 - 11 V, the anodization rate increases sharply (peak 1). The weak minimum observed at 13.5 V is attributed to the Al$_2$O$_3$/Al interface and a second maximum (peak 2) indicates an excess Al layer not oxidized during barrier formation. The smeared and weak minimum between 16 V and 18 V (see inset of **Figure 6**) is attributed to the Al/Nb interface and reflects the wetting effect of sputtered Al on polycrystalline Nb films. [34, 35]

The anodization spectrum of the memristive device Nb/Nb$_x$O$_y$/Al$_2$O$_3$-Al/Nb (green curve) shows all the above-mentioned features, except for a significantly larger peak 1. This can be explained by the high resistance of the oxygen-rich niobium oxide layer, because the voltage necessary to drive the current has to increase rapidly. This significant increase and decrease in anodization rate makes it difficult to investigate the quality of the underlying Al$_2$O$_3$ tunneling barrier. Given the broadened shape of peak 1 and the resolution of the anodization spectroscopy, it is difficult to estimate the influence of the reactive sputtering of NbO$_y$ on the Al$_2$O$_3$ tunnel barrier. Nevertheless, the DBM shows the same transition from the aluminum to the niobium layer as the Josephson junction (around 15 - 17 V). This suggests that the aluminum layer is not completely oxidized. Furthermore, the lack of an additional Nb oxide peak suggests the metallic low-resistance character of the Al$_2$O$_3$-back electrode interface.

To understand how an oxidized aluminum layer would influence the anodization spectrum, a further aluminum oxide layer was introduced: After one third of the regular Al thickness, the deposition was stopped and 1 mbar of oxygen was introduced (i.e. 1% of the pressure of the actual $Al_2O_3$ tunnel barrier). Afterwards, the remaining two-thirds of the aluminum layer was sputtered and oxidized as above, yielding the $Nb/Al_2O_3$-$Al/Al_2O_3$-$Al/Nb$ multilayer (red curve). The red curve in **Figure 6** shows the significant change compared to the Josephson junction and the DBM. This indicates that even the Al oxidized at 1 mbar is observable using anodization spectroscopy.

Anodization spectroscopy allows qualitative analysis of layer oxidation. To investigate the influence of the reactive sputtering on the electrical properties of the bottom electrode, samples with the following layer sequences were fabricated: 1) $Al_2O_3$-$Al/Nb$ and 2) $Nb_xO_y/Al_2O_3$-$Al/Nb$. The thickness of the bottom Nb electrode was reduced to approximately 4 nm, to emphasize the impact of the reactive sputtering. Afterwards, the resistance of both thin films was measured. If the reactive deposition is assumed to have a significant influence on the bottom electrode, this should lead to a drastic increase in film resistance, because the total multilayer thickness is comparable to the thickness of the oxidized layers observed in the STEM-EELS measurements. Both layers, however, showed ohmic behavior and had an almost identical resistance: Sample 1 had a resistance of (636 +/- 50) kΩ while sample 2 had a resistance of (635 +/-43) kΩ. This further confirms the metallic nature of the bottom electrode and is congruent to the observation of metallic character Nb suboxide in TEM investigations.

### 3.3. Kinetic Monte Carlo simulations

Within presented simulation approaches up to now, the resistances of the electrodes has been neglected. Models based on this device structure were able to explain important physical processes during the resistance change of the DBM device. [18] Although experimental results are not unambiguous, nano-spectroscopic analysis indicates the Nb bottom electrode to be partially oxidized (**Figure 2**) and to be most likely a Nb(II) oxide which is known to be almost metallic.[31] Comparative measurements have been done, using an $Al_2O_3$-Al/Nb and a $Nb_xO_y$/$Al_2O_3$-Al/Nb stack. The additional $Nb_xO_y$ turned out not to influence the stack resistance substantially. However, although this is a strong indicator, it does not prove a possible oxidation of the Nb electrode does not influence the overall device behavior of the DBM device. To further investigate the influence of a partially oxidized Nb electrode, a lumped element circuit model has been set up and solved with the simulation tool SPICE. **Figure 7 (a)** shows the simulated circuit. Here D1 represents the Au/$NbO_y$ Schottky contact. The underlying equation, describing the current through the Schottky contact, following the thermionic emission theory, is given by:

$$I_S = I_R \cdot \exp\left(\frac{eV}{nk_BT} - 1\right) \qquad (2)$$

$$I_R(V > 0) = A \cdot AT^2 \cdot \exp\left(\frac{-\Phi_b}{k_BT}\right) \qquad (3)$$

$$I_R(V < 0) = -A \cdot AT^2 \cdot \exp\left(\frac{-\Phi_b}{k_BT}\right) \exp\left(\frac{\alpha_r\sqrt{|V|}}{k_BT}\right) \qquad (4)$$

where $\Phi_b$ is the Schottky barrier height (0.83 eV), $e$ is the elementary charge, $A$ is the junction area, $k_B$ and $T$ are the Boltzmann constant and the temperature, $A^*$ is the effective Richardson constant ($1.20173 * 10^6\ A/m^2K^2$), $n$ is the ideality factor (3.6), $\alpha_r$ denotes a device dependent parameter (0.15) and $V$ is the voltage drop over the Schottky contact. The parameters of the circuit have been chosen so that the device is in the low resistance state as presented in [18]. The resistance $R_t$ (t stands for the top Nb oxide layer) represents the $NbO_y$ layer. The resistance is calculated by:

$$R_t = \frac{\rho_t l_t}{A} \qquad (5)$$

Where $l_t$ is the thickness and $\rho_t$ is the specific resistance ($2.5 \cdot 10^3\,\Omega\text{m}$) of the NbO$_y$ layer. The varistor VDR represents the AlO$_x$ tunnel barrier. The current through the tunnel barrier is given by the current formula of Simmons:

$$I_T = \frac{Ae}{2\pi hd} \cdot \left( x_1 \cdot \exp\left[-\frac{4\pi d\sqrt{2m}}{h}\sqrt{x_1}\right] - x_2 \cdot e \exp\left[-\frac{4\pi d\sqrt{2m}}{h}\sqrt{x_2}\right] \right) \quad (6)$$

with $x_1 = e\Phi_0 - eV/2$ and $x_2 = e\Phi_0 + eV/2$. $\Phi_0$ is the barrier height of the tunnel barrier (3.1 eV), $V$ is the voltage drop over the tunnel barrier, $m$ is the electron mass, $h$ is Planck's constant and $d$ is the thickness of the tunnel barrier. Up to now, this model is very similar to the electrical model presented in [16]. Within this work, this model has been extended by a second resistance $R_b$ (b for bottom), representing the partially oxidized Nb electrode. This resistance is calculated by:

$$R_b = \frac{\rho_b l_b}{A} \quad (7)$$

Here $l_b$ is the thickness of the oxidized part of the Nb electrode (measured to be about 6.5 nm) and $\rho_b$ is the specific resistance of the NbO$_z$ layer.

The voltage drop over the NbO$_y$ part (represented by $R_t$) gives deeper insights into the behaviour of the global device. At first, the applied voltage drop leads to an electric field within the NbO$_y$ layer. This electric field itself leads to a motion of oxygen vacancies and thus to a resistance change of the device as presented in [18]. This means, the higher the voltage drop over the NbO$_y$ part, the higher the resulting electric field and the higher the resistance change. Hence, the voltage drop provides information about the resistive switching. Furthermore, due to the series like circuit, the current through the device is proportional to the voltage drop over the NbO$_y$ part. Hence, the voltage drop over the NbO$_y$ part also provides information about the device current. The resistance $R_b$ influences the voltage drop over the resistance $R_t$ and therefore over the NbO$_y$ part. Therefore, at first the voltage drop over $R_t$ is calculated without $R_b$ for two different voltages applied to the overall device (2 V and 3 V).

Afterwards, the voltage drop over $R_t$ is calculated including $R_b$ to the circuit also for 2 V and 3 V respectively. The change of the voltage drop over $R_t$ is a measure for the influence of the $NbO_z$ part. Since the $NbO_z$ layer was measured to be Nb(II) oxide, the specific resistance $\rho_b$ is expected to be on the order of $10^{-6}$ Ωm. [33] However, the exact specific resistance is not known. Therefore, the voltage drop over $R_t$ is calculated for different values of $R_b$, resulting from different $\rho_b$ ranging from $10^{-6}$ Ωm up to $10^4$ Ωm.

**Figure 7 (c)** shows the relative voltage drop $V/V_0$ over the $NbO_y$ layer for different specific resistances $\rho_b$ of $R_b$. As a result, up to a specific resistance of $10^2$ Ωm, which is already the specific resistance of Nb(IV) oxide, there is nearly no change in the voltage drop over the NbOy layer. For higher specific resistances the relative voltage drop changes strongly, indicating a large influence of $R_b$ on the switching mechanism. Smaller applied voltages lead to a lower change in the relative voltage. This is shown for an example voltage of 2 V. As long as the change in relative voltage drop is small, the voltage drop over the NbOz layer is negligible. This results in a small electric field. Under this condition, the motion of defects within this layer can be neglected, due to the negligible electric field. The simulation results underline that the partially oxidized Nb electrode influences the overall behaviour of the device only marginally.

### 4. DISCUSSION

As apparent from **Figure 2** the interfaces around the top $NbO_y$ layer are extremely smooth with virtually no detectable surface roughness. This is extremely important for the proposed switching mechanism, as it ascertains homogenous, undistorted electrical fields, in turn allowing for homogenous memristive switching. Furthermore, the investigations underline the high oxidation state of the $NbO_y$ layer. Being at least Nb(IV) oxide, this ensures a functioning Schottky barrier between the Au top electrode and $NbO_y$ especially in combination with the layer thickness, closely corresponding to the targeted values.

As for the oxidized bottom electrode, sub-oxides of Nb are known to be either n-type semiconductors or even metallic conductors. Thus, the existence of pronounced Nb BE oxidation does not hamper the memristive mechanism proposed by Hansen et al, as also underlined by the electrical simulations and anodization spectroscopy. The reason for oxidation of the BE cannot be resolved with certainty. Given that Al is known to exhibit self-limiting oxidation [37] at room temperature and standard atmosphere of around only 1 nm, it seems plausible that another external driving force is required to oxidize the Al completely. Observation of this self-limited oxidation in the Josephson junction (cf. supplement) suggests that through-oxidation of the Al and the oxidation of the BE is linked to the additional deposition steps, namely the reactive deposition of $NbO_y$. Here, the Ar/O plasma ignited for reactive deposition can serve as a source for oxidation of the underlying layers, the driving force being a combination of concentration gradient from high (plasma) to low ($AlO_x$ and at the time of deposition the metallic Nb BE) and the kinetic energy and charge of the oxygen ions. It should be noted that despite this difference in targeted and actual layer sequence the device still exhibits memristive properties as described by Hansen et al, underlining the superior robustness to deviations of the DBM approach. Anodization spectroscopy data further substantiate this by proving the metallic character of the bottom electrode and underlining that reactive sputtering above a metallic Nb electrode has no significant influence on electrical behaviour.

The finding of O vacancies in the $AlO_x$ tunnel barrier suggests that this thought to be passive layer might change its resistivity during bias application, after all. However, the number of vacancies is expected to be low enough to not have an influence on tunnelling behaviour. In particular, the amorphous nature of the tunnel barrier implies that these defects have little to no effect as discussed by Pesenson et al. [38] The lowest calculated defect concentration of 1.7 % results in a theoretical average inter-defect distance of 29.5 Å. Considering the tunnelling thickness of 1 nm, this inter-defect distance is expected to be large enough to not have an influence. Potential existence of O vacancies in either of the Nb oxide layers cannot be measured in the same way. Detection of oxygen vacancies in $AlO_x$ relies on their effect of inter-band states. Only if O vacancies in Nb oxide had the same effect would they be detectable in a similar fashion.

The discrepancy of potentially finding metallic Al during anodization spectroscopy and its absence in STEM investigations can be caused by different factors. Anodization spectroscopy was conducted with large scale samples; hence the result could imply that TEM samples were taken from randomly oxidized regions. However, during the course of investigations around ten lamellae were extracted from different DBM devices and none of them showed metallic Al, indeed all samples exhibit congruent results. On the other hand, anodization spectroscopy relies on the change of resistivity characteristic for different materials. Yet, no complete theoretical description or the possibility of simulation for these spectra exist, potentially hampering correct interpretation. The surface roughness between the Al und Nb oxides might simply render the thicker $AlO_x$ regions undetectable for anodization spectroscopy and the effect of a metallically-conducting Nb suboxide is also unknown.

## 5. CONCLUSION

Intensive spectroscopic investigations on memristive $AlO_x/NbO_y$ devices and their Josephson junction base were conducted.

1. Slight oxidation of the BE was found. As the layer sequence of the Josephson junction base has been verified beforehand, oxidation possibly took place in the subsequent deposition of $NbO_y$ on top of the junction's $AlO_x$ tunnel barrier. Simulations in accordance with EELS and anodization spectroscopy underline the low resistance of this layer.

2. Oxidation due to TEM preparation artifacts has been excluded by a comprehensive approach including low energy Argon ion thinning by PIPS, standard FIB lamella thinning, and backside FIB lamella thinning.
3. An unexpected content of oxygen vacancies in the $AlO_x$ was found its effect on memristive behaviour is unknown. It was not possible to test for oxygen vacancies in Nb oxide in the same manner.
4. Nano-spectroscopic investigations showed that surface roughness limits the thickness of the tunnel barrier to around 1 nm and the Schottky barrier forming $NbO_y$ to around 2 nm, thus closely corresponding to the targeted layer design. Importantly, surface roughness around the Schottky barrier forming layers is virtually zero, thus ensuring a homogenous electrical field distribution and in turn homogenous memristive switching.

Overall, these results demonstrate the high robustness of the DBM regarding deviations from the targeted design and layer sequence. However, they also open up the path towards potential improvements.

- Deposition of a thicker Al layer could inhibit the oxidation of the bottom electrode.
- An additional adhesion layer might reduce the surface roughness visible between the (oxidized) Nb bottom electrode and the Al layer. Alternatively, deposition of Al oxide by atomic layer deposition has been shown to produce atomically smooths layers.
- The latter could also resolve the presence of oxygen vacancies in the tunnel barrier.

6. ACKNOWLEDGEMENTS

The authors gratefully acknowledge funding by the DFG through the research unit FOR2093 and by the European Union within the 7th Framework Program (FP7/2007-2013) under Grant Agreement no. 312483 (ESTEEM2). JS and LK would like to thank Mrs. Christin Szillus and Mrs. Martina Dienstleder for sample preparation by FIB and NanoMill and Prof. Ferdinand Hofer as well as Prof. Dagmar Gerthsen for supporting the investigations. The authors acknowledge the support of the Karlsruhe Nano Micro Facility (KNMF) for Electron Microscopy and Spectroscopy facilities.

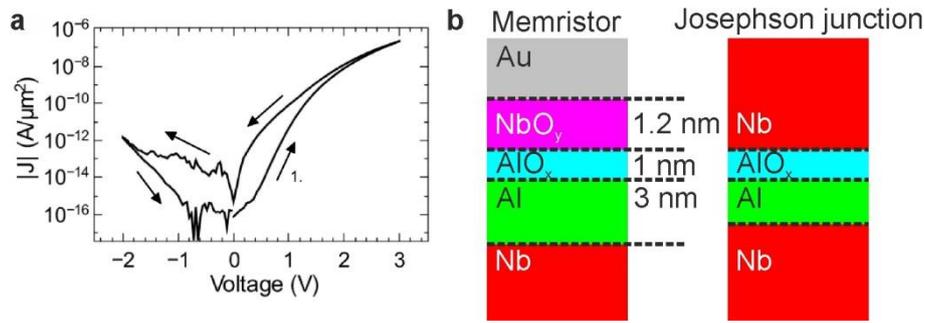

**Figure 1:** The DBM device as presented by Hansen et al (2015). (a) Measured I-V curve with distinct low- and high-resistance states. (b) Layer scheme and approximate thicknesses expected from deposition parameters for the memristor. (c) Design-wise similar Josephson junction which was analysed in this study because of its comparability to the DBM.

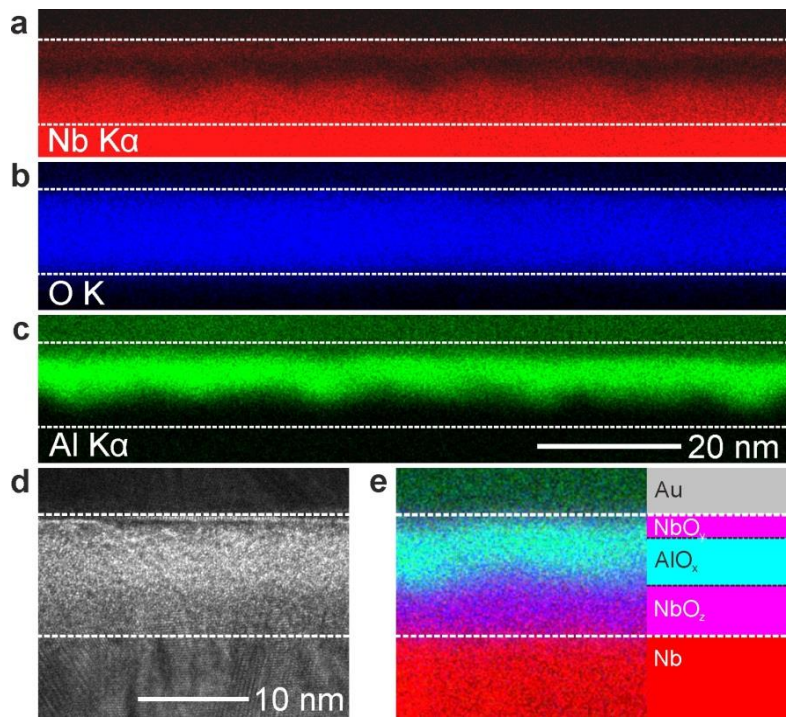

**Figure 2:** Structural and chemical analysis of the DBM. (a) to (c) show STEM-EDX maps of Nb, O and Al respectively taken from a backside thinned sample. To enhance visibility of the $NbO_y$ layer, intensity, contrast and brightness in (a) have been modified. Dashed white lines indicate the extent of the oxide layers. (d) HRTEM micrograph showing the crystalline nature of the Au and Nb electrodes and amorphous character of the oxide layers between. (e) Compound map created from maps (a) to (c) proving that the amorphous region in (d)

corresponds to the NbO$_y$/AlO$_x$/NbO$_z$ layers next to a sketch (not to scale) of the observed layer sequence as opposed to the sequence shown in **Figure 1**.

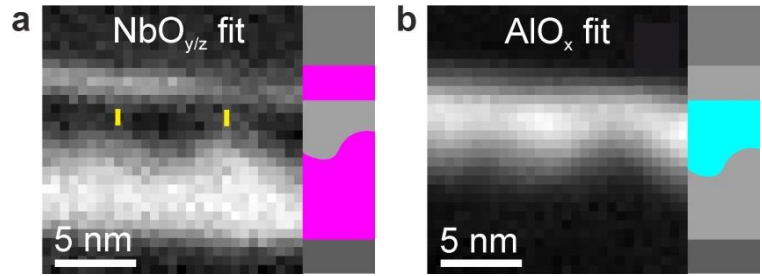

**Figure 3:** Goodness of fit map of the O-K edge with an O-K edge reference of (a) NbO$_{y/z}$ and (b) AlO$_x$. The references with which the fit was performed were extracted from data of the same set and references from several samples are displayed in **Figure S1**. While the EDX maps in **Figure 2** and the AlOx fit in (b) suggest that the AlOx is around 5 nm thick, the Nb oxide fit in (a) proves that surface roughness effects limits the tunnel barrier thickness to around 1 nm at the thinnest spots. The yellow lines are 1 nm and indicate the thinnest locations in the tunnel barrier. For a detailed description of data handling cf. section II.

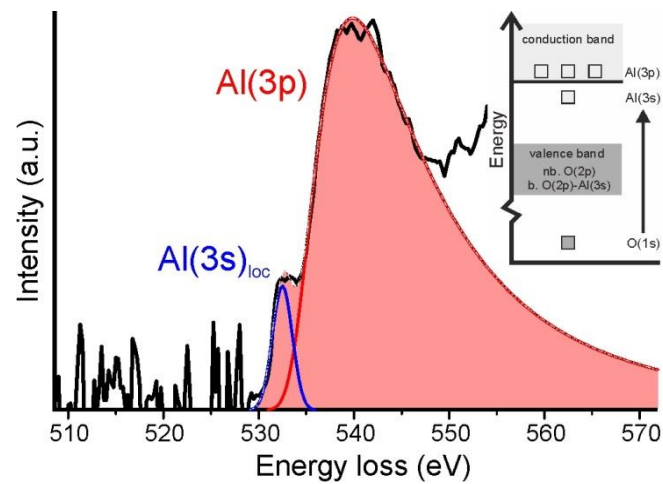

**Figure 4:** Close up of the O-K edge of the AlO$_x$ region including the fit line. Spectrum has undergone background subtraction prior to fitting. A Gaussian peak (blue) was chosen for the satellite feature and a Pearson IV function (red) was fit to the main edge of oxygen, the combined fit is indicated by the red area. The ratio of cross sections is 4.3 %. Inset shows a reduced diagram of the excitation process from O(1s) to the Al states above the Fermi energy.

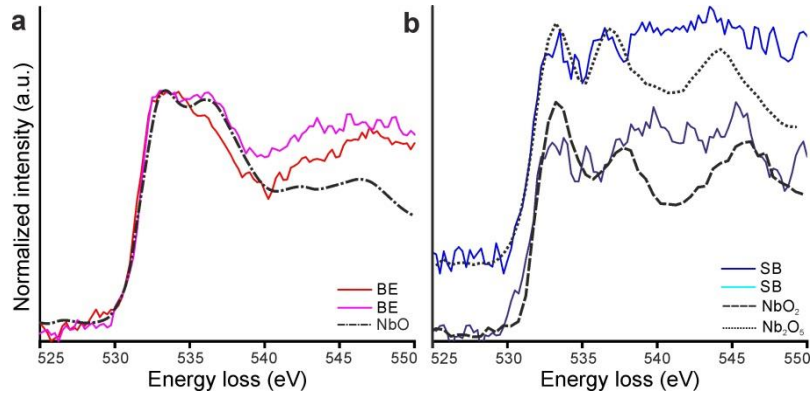

**Figure 5**: EEL spectra containing the O K edge of different samples, extracted from Nb oxide containing regions. a) The shape of the $NbO_y$ bottom electrode ELNES matches the fingerprints of Nb(II) oxide. b) The Schottky barrier oxide is most likely a mixture of oxidation states Nb(IV) and Nb(V). The net intensity deviation a higher energy losses is likely due to incorrect background subtraction and plural scattering effects.

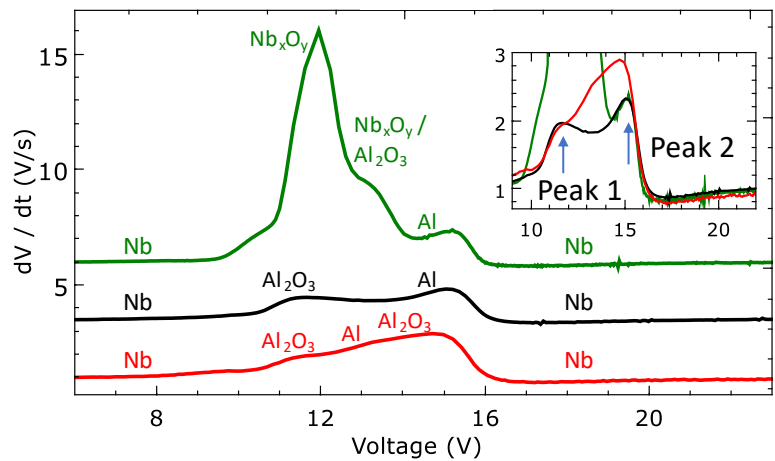

**Figure 6**: Anodization spectra of a Josephson junction (black curve), the DBM (green curve) and a sample to show the influence on the anodization spectroscopy of a partially oxidized aluminum layer (red curve). All curves were normalized to an anodization rate of 1 V/s for niobium. The spectra were vertically shifted for a better illustration. The inset shows all curves on top of each other, to visualize the different anodization rates around peak 2.

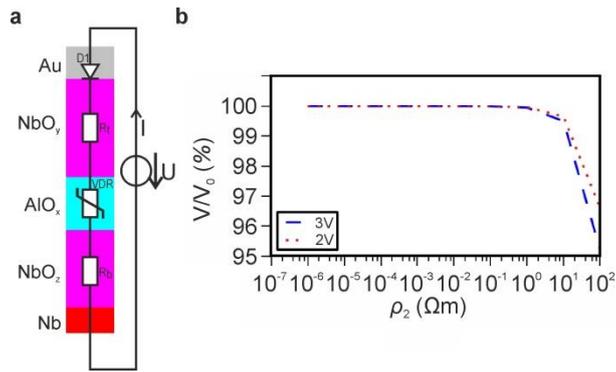

**Figure 7:** Input and results for kinetic Monte Carlo simulations. (a) Lumped element model of the DBM device. D1 represents the Schottky contact, $R_t$ the $NbO_y$ part, VDR the $Al_2O_3$ tunnel barrier and $R_b$ the oxidized Nb electrode. (b) Voltage drop V over the $NbO_y$ part including $R_b$ in terms of the voltage drop $V_0$ over the $NbO_y$ part without $R_b$ for different specific resistances $\rho_b$ of $R_b$..